\documentclass[preprint]{aastex}

\shorttitle{Gravitational Lensing by Nearby Clusters of Galaxies}
\shortauthors{Cypriano et al.}

\newcommand\eq{\begin{equation}}
\newcommand\eeq{\end{equation}}
\newcommand\eqn{\begin{eqnarray}}
\newcommand\eeqn{\end{eqnarray}}
\newbox\grsign \setbox\grsign=\hbox{$>$} \newdimen\grdimen
\grdimen=\ht\grsign
\newbox\simlessbox \newbox\simgreatbox
\setbox\simgreatbox=\hbox{\raise.5ex\hbox{$>$}\llap
     {\lower.5ex\hbox{$\sim$}}}\ht1=\grdimen\dp1=0pt
\setbox\simlessbox=\hbox{\raise.5ex\hbox{$<$}\llap
     {\lower.5ex\hbox{$\sim$}}}\ht2=\grdimen\dp2=0pt

\newcommand\simless{\mathrel{\copy\simlessbox}}
\newbox\simppropto
\setbox\simppropto=\hbox{\raise.5ex\hbox{$\sim$}\llap
     {\lower.5ex\hbox{$\propto$}}}\ht2=\grdimen\dp2=0pt

\input{epsf.sty}

\begin{document}

\title{Gravitational Lensing by Nearby Clusters of Galaxies}

\author{Eduardo S. Cypriano and Laerte Sodr\'e Jr.}
\affil{Departamento de Astronomia, Instituto Astron\^omico e Geof\'{\i}sico
da USP, Av. Miguel Stefano 4200, 04301-904 S\~ao Paulo, Brazil}
\email{eduardo,laerte@iagusp.usp.br}

\author{Luis E. Campusano \altaffilmark{1}}
\affil{Observatorio Astron\'omico Cerro Cal\'an, Departamento de 
Astronom\'{\i}a, Universidad de Chile, Casilla 36-D, Santiago, Chile}
\email{lcampusa@das.uchile.cl}

\author{Jean-Paul Kneib}
\affil{Observatoire Midi-Pyr\'en\'ees, Laboratoire d'Astrophysique,
UMR 5572, 14 Avenue E. Belin, 31400 Toulouse, France}
\email{kneib@ast.obs-mip.fr}

\author{Riccardo Giovanelli, Martha P. Haynes and Daniel A. Dale \altaffilmark{1}}
\affil{Center for Radiophysics and Space Research and National Astronomy and
Ionosphere Center, Cornell University, Ithaca, NY 14853}
\email{riccardo,haynes,dale@astrosun.tn.cornell.edu}

\and
\author{Eduardo Hardy \altaffilmark{1}}
\affil{National Radio Astronomy Observatory, Santiago, Chile}
\email{ehardy@nrao.edu}

\altaffiltext{1}{Visiting Astronomer at Cerro Tololo Inter-American
Observatory. CTIO is operated by AURA, Inc.\ under contract to the National Science
Foundation.}

\begin{abstract}
We present an estimation of the expected number of arcs and arclets in a sample
of nearby ($z < 0.1$) clusters of galaxies, that takes into account the
magnitude limit of the objects as well as seeing effects. We show that strong
lensing effects are not common, but also they are not as rare as usually 
stated. Indeed, for a given cluster, they present
a strong dependence with the magnitude limit adopted in the analysis
and the seeing of the observations.
We also describe the procedures and results
of a search for lensing effects in a sample of 33 clusters spanning the
redshift range of 0.014 to 0.076, 
representative of the local cluster distribution. This search produced two
arc candidates. The first one is in A3408 ($z=0.042$) , the same arc previously
 discovered by \citet{CeH}, with $z = $ 0.073 and associated to the brightest
cluster galaxy. The second candidate is in the cluster A3266 ($z=0.059$)
and is near a bright elliptical outside the cluster center, requiring the
presence of a very massive sub-structure around this galaxy to be produced by
gravitational lensing.
\end{abstract}


\section{Introduction}
Gravitational lensing is a powerful technique to probe distant galaxies
as well as for studying  the matter distribution in galaxy clusters. 
Indeed, the analysis of bright
arcs and arclets  or other distortions (weak lensing), induced by 
the gravitational lensing of background sources by a galaxy cluster, 
has allowed the 
determination of the mass distribution in these structures, independently of
other more common techniques, like the application of the virial theorem 
or the analysis of the X-ray emission.

Most studies of gravitational lensing by clusters have been focused
in distant objects, resulting in the discovery of rather high $z$
arcs and arclets. 
Galaxy lensing by clusters, assuming a non-evolving mass profile and a
reasonable redshift distribution for the faint galaxy population, has its 
maximum around $z\sim 0.2$ \citep[e.g.][]{Nat&Keneib}. However, this
does not imply that the lensing efficiency of nearby clusters is totally 
negligible and, in fact, several groups have recently found evidence of
strong lensing effects in low-redshift clusters. For instance,
\citet{AFK96} discovered a $z=0.43$ arc in the cluster PKS0745-191
(at $z=0.103$), that has been successfully modeled as a gravitational
lens image. \citet{SBH96} found an arc-like structure,
still without redshift information, near NGC4881 in the Coma cluster 
($z=0.024$). \citet{CeH}  found an arc-like object at 
$z=0.073$ in A3408, a cluster at $z=0.042$. Lens models of this structure
are discussed by \citet{CKH98}. \citet{B&M99} 
discovered an arc-like object in A2124 ($z=$0.066) that is
probably the lensed image of a galaxy at $z$=0.573. \citet{CKH98}
 predicted the detection of weak-shear in low ($z<0.1$) clusters, 
which has been recently confirmed by \citet{Jetal99} 
and Kneib et al. (2000, in preparation).

It is worth to point out that the
study of gravitational lensing by low redshift clusters presents, in
principle, an important advantage when compared with those of 
more distant clusters: due to the large angular diameter that nearby 
clusters have, gravitational lensing
may allow examining in great spatial detail the mass distribution of their
central regions. Although the lensing efficiency of a cluster
depends strongly on its central mass distribution, the latter is usually not
well known. Observations of lensing effects allow probing low redshift clusters
with high spatial resolution and, consequently, they can help to add new
constraints on the mass distribution in the centers of these structures.
For the weak lensing regime \citep[see][for a review]{Mellier99}, the magnitude 
of the effects in high
redshift clusters depends strongly on the imprecisely known
redshift distribution of the faint background galaxies, while in 
low-$z$ clusters the weak lensing effects are almost independent of it.

In this paper we present a simple estimation of the expected number of 
arcs and arclets in low redshift clusters, as well as  the results of
an analysis of a sample of nearby clusters ($z \le 0.076$), where we have
looked for arc-like structures that may be produced by gravitational
lensing, either by the central cluster potential as a whole or by
substructures in the mass distribution related to a galaxy
not located at the fiducial center of the cluster.
Following \citet{HKM99} we call an arc a structure distorted
by gravitational lensing with axial ratio (length-to-width) larger than 10, 
and arclet a structure with axial ratio smaller than 10.

The layout of the paper is as follows.
The sample of galaxy clusters analyzed here is discussed in Section 2. 
The estimation of the number of luminous arcs and arclets
expected in a sample of clusters is presented in Section 3 (and Appendix A).
In Section 4 we describe the search for arcs and arclets in the sample.
The two arclet-like structures found in our 
search are presented and discussed in Section 5. Follow up observations
of these arclets are also presented in this section.
Finally, we summarize our conclusions in Section 6. When necessary,
we adopt $H_0 = 50$ h$^{-1}_{50}$ km s$^{-1}$ Mpc$^{-1}$ and, unless where
explicitly stated, $\Omega_0$ = 1 and $\lambda$ = 0.

\section{The sample of nearby galaxy clusters }
The images analyzed in this project were originally obtained as part of 
the PhD  thesis of Daniel A. Dale on peculiar motions of clusters with 
$z < 0.1$ \citep{Dale97,Dale98,Dale99a,Dale99b,Dale99c}. The sample of galaxy 
clusters 
with some relevant properties is presented in Table \ref{sample}. 

\renewcommand{\arraystretch}{.5}
\begin{deluxetable}{lcccccccc}
\tablecaption{Cluster sample - Col. 1:
name of the cluster in the Abell catalog (Abell, Corwin \& Olowin 1989); 
cols. 2 and 3: equatorial coordinates (J2000) of the cluster 
centers;
col. 4: redshift; 
col. 5: radial velocity dispersion;
col. 6: richness class (Abell, Corwin \& Olowin 1989); 
col. 7: X-ray luminosity (0.1-2.4 Kev) from Ebeling et al. (1996);
col. 8: number of images per cluster considered in this study;
col. 9: expected number of arcs in OCDM 
cosmology (see Section 3 and Appendix A) assuming the same seeing 
of 1.4$^{\prime\prime}$ (FWHM) for all images.
\label{sample}}
\tablehead{
\colhead{Name} & \colhead{$\alpha_{2000}$} & \colhead{$\delta_{2000}$} & \colhead{z} & \colhead{$\sigma_v$} &
\colhead{R} & \colhead{L$_X$} & \colhead{Num} & \colhead{$<$N$>$} \\

\colhead{} & \colhead{} & \colhead{} & \colhead{} & \colhead{(km s$^{-1}$)} & \colhead{} & 
\colhead{(10$^{44}$ erg s$^{-1}$)} & \colhead{} & \colhead{($\times 1000$)} \\

\colhead{(1)} & \colhead{(2)} & \colhead{(3)} & \colhead{(4)} & \colhead{(5)} & \colhead{(6)} & 
\colhead{(7)} & \colhead{8} & \colhead{(9)}  
}
\startdata

A85   &~0$^h$ 41$^m$ 37$^s$ &~-9$^\circ$  20$^\prime$.6 & 0.0559 \tablenotemark{a} & ~969 \tablenotemark{d} & 1 & 8.38    &~1 & 22.7\\
A114  &~0$^h$ 53$^m$ 39$^s$ &-21$^\circ$  40$^\prime$.7 & 0.0572 \tablenotemark{b} & ~904 \tablenotemark{b} & 0 & $<$ 0.7 &19 & 16.9\\
A119  &~0$^h$ 56$^m$ 21$^s$ &~-1$^\circ$  15$^\prime$.8 & 0.0438 \tablenotemark{b} & ~721 \tablenotemark{b} & 1 & 3.23    &~1 & 8.2\\
A194  &~1$^h$ 25$^m$ 33$^s$ &~-1$^\circ$  30$^\prime$.4 & 0.0168 \tablenotemark{b} & ~409 \tablenotemark{b} & 0 & 0.14    &~1 & 1.3\\
A295  &~2$^h$ ~2$^m$ 27$^s$ &~-1$^\circ$  ~4$^\prime$.6 & 0.0424 \tablenotemark{a} & ~359 \tablenotemark{d} & 1 & $<$ 0.4 &~2 & 0.5\\
A496  &~4$^h$ 33$^m$ 37$^s$ &-13$^\circ$  14$^\prime$.8 & 0.0327 \tablenotemark{b} & ~715 \tablenotemark{b} & 1 & 3.54    &18 & 9.4\\
A548  &~5$^h$ 47$^m$ ~4$^s$ &-25$^\circ$  37$^\prime$.6 & 0.0415 \tablenotemark{a} & ~576 \tablenotemark{d} & 2 & 0.30    &~9 & 3.5\\
A1736 &13$^h$ 26$^m$ 46$^s$ &-27$^\circ$  ~6$^\prime$.6 & 0.0357 \tablenotemark{b} & ~390 \tablenotemark{b} & 0 & 2.37    &10 & 0.8\\
A2670 &23$^h$ 54$^m$ 10$^s$ &-10$^\circ$  24$^\prime$.3 & 0.0759 \tablenotemark{a} & ~852 \tablenotemark{d} & 3 & 2.55    &~2 & 10.4\\
A2806 &~0$^h$ 40$^m$ 10$^s$ &-56$^\circ$  ~9$^\prime$.5 & 0.0262 \tablenotemark{c} & ~390 \tablenotemark{b} & 0 & $<$ 0.2 &16 & 1.1 \\
A2870 &~1$^h$ ~7$^m$ 43$^s$ &-46$^\circ$  55$^\prime$.0 & 0.0250 \tablenotemark{c} & \nodata                & 0 & $<$ 0.1 &10 & \nodata \\
A2877 &~1$^h$ ~9$^m$ 56$^s$ &-45$^\circ$  55$^\prime$.9 & 0.0248 \tablenotemark{a} & ~887 \tablenotemark{d} & 0 & 0.46    &16 & 25.1\\
A2911 &~1$^h$ 26$^m$ 12$^s$ &-37$^\circ$  56$^\prime$.4 & 0.0202 \tablenotemark{c} & ~547 \tablenotemark{d} & 1 & $<$ 0.1 &10 & 3.9\\
A3193 &~3$^h$ 58$^m$ 13$^s$ &-52$^\circ$  20$^\prime$.5 & 0.0351 \tablenotemark{b} & ~624 \tablenotemark{b} & 0 & $<$ 0.3 &10 & 5.2  \\
A3266 &~4$^h$ 31$^m$ 10$^s$ &-61$^\circ$  26$^\prime$.7 & 0.0593 \tablenotemark{b} & 1103 \tablenotemark{b} & 2 & 6.15    &~4 & 36.3\\
A3367 &~5$^h$ 49$^m$ 21$^s$ &-24$^\circ$  28$^\prime$.1 & 0.0426 \tablenotemark{c} & \nodata                & 0 & $<$ 0.4 &~1 & \nodata   \\
A3376 &~6$^h$ ~0$^m$ 43$^s$ &-40$^\circ$  ~3$^\prime$.0 & 0.0490 \tablenotemark{a} & ~688 \tablenotemark{d} & 0 & 2.48    &~2 & 6.4\\
A3381 &~6$^h$ ~9$^m$ 54$^s$ &-33$^\circ$  36$^\prime$.0 & 0.0384 \tablenotemark{b} & ~304 \tablenotemark{b} & 1 & $<$ 0.2 &~6 & 0.3\\
A3389 &~6$^h$ 21$^m$ 47$^s$ &-64$^\circ$  57$^\prime$.6 & 0.0267 \tablenotemark{c} & ~598 \tablenotemark{d} & 0 & 0.32    &~3 & 5.0\\
A3395 &~6$^h$ 27$^m$ 31$^s$ &-54$^\circ$  24$^\prime$.0 & 0.0506 \tablenotemark{a} & ~852 \tablenotemark{d} & 1 & 2.80    &~2 & 14.6\\
A3407 &~7$^h$ ~5$^m$ ~1$^s$ &-49$^\circ$  ~4$^\prime$.6 & 0.0429 \tablenotemark{b} & ~490 \tablenotemark{b} & 1 & $<$ 0.4 &~4 & 1.8 \\
A3408 &~7$^h$ ~8$^m$ 31$^s$ &-49$^\circ$  12$^\prime$.9 & 0.0420 \tablenotemark{c} & ~900 \tablenotemark{e} & 0 & 0.50   &~6 & 20.6\\
A3528 &12$^h$ 54$^m$ 18$^s$ &-29$^\circ$  ~1$^\prime$.3 & 0.0530 \tablenotemark{c} & ~972 \tablenotemark{d} & 1 & 1.33    &~8 & 8.0\\
A3558 &13$^h$ 28$^m$ 00$^s$ &-31$^\circ$  30$^\prime$.5 & 0.0488 \tablenotemark{b} & ~965 \tablenotemark{b} & 4 & 6.27    &15 & 24.6\\
A3564 &13$^h$ 34$^m$ 22$^s$ &-35$^\circ$  13$^\prime$.5 & 0.0491 \tablenotemark{c} & \nodata                & 1 & $<$ 0.5 &~7 & \nodata   \\
A3571 &13$^h$ 47$^m$ 28$^s$ &-32$^\circ$  52$^\prime$.1 & 0.0396 \tablenotemark{c} & 1045 \tablenotemark{d} & 2 & 7.36    &~2 & 38.7\\
A3572 &13$^h$ 48$^m$ 11$^s$ &-33$^\circ$  22$^\prime$.9 & 0.0405 \tablenotemark{c} & \nodata                & 0 & $<$ 0.4 &~4 & \nodata  \\
A3574 &13$^h$ 49$^m$ ~9$^s$ &-30$^\circ$  17$^\prime$.9 & 0.0141 \tablenotemark{c} & ~491 \tablenotemark{d} & 0 & $<$ 0.1 &~5 & 2.8\\
A3656 &20$^h$ ~0$^m$ 32$^s$ &-38$^\circ$  31$^\prime$.7 & 0.0186 \tablenotemark{b} & ~350 \tablenotemark{b} & 0 & $<$ 0.1 &22 & 0.7\\
A3667 &20$^h$ 12$^m$ 30$^s$ &-56$^\circ$  49$^\prime$.0 & 0.0550 \tablenotemark{b} & ~1059 \tablenotemark{b}& 2 & 8.76    &22 & 32.8\\
A3716 &20$^h$ 51$^m$ 30$^s$ &-52$^\circ$  43$^\prime$.0 & 0.0454 \tablenotemark{b} & ~842 \tablenotemark{b} & 1 & 1.01    &26 & 15.0\\
A3744 &21$^h$ ~7$^m$ 12$^s$ &-25$^\circ$  29$^\prime$.0 & 0.0371 \tablenotemark{b} & ~570 \tablenotemark{b} & 1 & $<$ 0.3 &30 & 3.6\\
A4038 &23$^h$ 47$^m$ 54$^s$ &-28$^\circ$  09$^\prime$.3 & 0.0291 \tablenotemark{b} & ~843 \tablenotemark{b} & 2 & 1.90    &12 & 19.2\\
\enddata
\tablenotetext{a}{Wu, Fand \& Xu (1998)}
\tablenotetext{b}{Dale et al. (1999c)}
\tablenotetext{c}{Abell, Corwin \& Olowin (1989)}
\tablenotetext{d}{Fadda et al. (1996)}
\tablenotetext{e}{Campusano, Kneib \& Hardy (1998)}\\
\tablecomments{in column 4 references a and b give heliocentric redshifts while in 
reference c the redshifts are relative to the CMBR.} 
\end{deluxetable}

The observational material that is analyzed here consists of several
 Kron-Cousins I-band images obtained with the 0.9m  CTIO telescope. 
The details of the observations are discussed in \citep{Dale97,Dale98}
and here we only summarize them. The detector used was the 2k$\times$2k
Tek2k No.3 CCD, with a scale of 0.396  arcsec per pixel, 
resulting in a field of 13.5$^\prime \times 13.5^\prime$ per image. 
The exposure times were of 600 seconds in all cases. The images reach $\sim$ 23.7 I 
mag arcsec$^{-2}$ at the 1.0 $\sigma$ level over the sky background. 
In Section 4 we
estimate that the  isophotal completeness limit for extended sources in these
 images is
$\sim$19.5 mag (1.5 $\sigma$ in 40 connected pixels), corresponding to
a total magnitude limit of $\sim$ 19.0 mag.
The average seeing of the images is 1.4$^{\prime \prime}$.
Although these images are not very deep, this material is the same where 
\citet{CeH} found an arc in A3408. 
These images in general do not uniformly cover a cluster, since they 
were taken in regions near spiral galaxies, but the central region
of the clusters are always covered. The number of images analyzed per
cluster are also presented in Table \ref{sample}.

\begin{figure}[h]
\epsfxsize=8cm
\centerline{\epsfbox{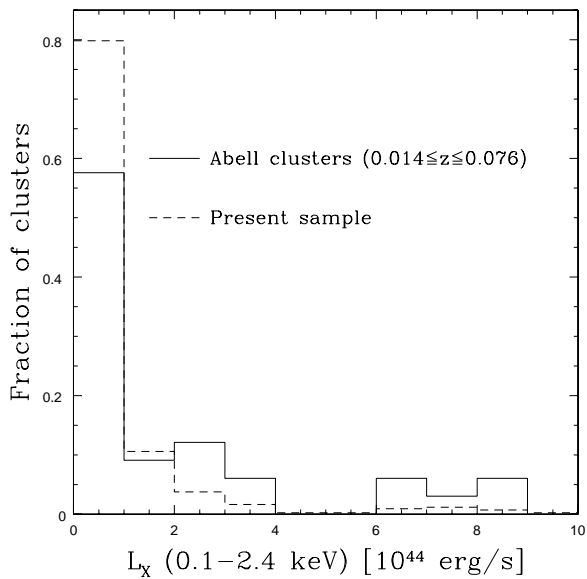}}
\vskip 0.5cm
\caption{Relative distribution of X-ray luminosities (0.1-2.4 Kev) of the 
present sample and the XBAC's sample
with measures done by the ROSAT all sky survey (Ebeling et al. 1996) and 
all Abell clusters in the same redshift range (solid line).}
\label{xcomp}
\end{figure}

This sample is a good representation of the cluster
distribution in the nearby universe. Its richness distribution  
is presented in Table \ref{rich}. It is 
consistent with the whole Abell catalog \citep{ACO} ,
with a slight excess of high richness clusters.
In Figure \ref{xcomp} we compare the X-ray luminosity distribution
of our sample with that of the XBAC catalogue \citep{XBACS}
in the same redshift range (i.e., 0.014 $\le z \le$ 0.076).
The XBAC catalogue is a X-ray flux limited 
catalog of Abell clusters from the ROSAT all sky survey.
Note that the clusters that were not detected by ROSAT have been
included in the first bin in Figure \ref{xcomp}.
This figure indicates that the X-ray luminosity distribution
of our sample presents an overall agreement with the cluster
distribution of the nearby Universe.

\placefigure{xcomp}

If richness and X-ray luminosity are proportional to  cluster mass, we
conclude that the  mass distribution of our cluster sample is
representative of that actually present at low redshifts.

\renewcommand{\arraystretch}{1.0}
\begin{deluxetable}{crr}
\tablecaption{Richness distribution of clusters in
the Abell catalog and in the present sample.
\label{rich}}
\tablehead{\colhead{Richness class} & \colhead{Abell catalog} &
\colhead{Present sample}} 
\startdata
0 & 60\%   & 42\% \\
1 & 30\%   & 36\% \\
2 & 9\%    & 15\% \\
3 & 2\%    & 3\%  \\
4 & 0.2\%  & 3\%  \\
5 & 0.02\% & 0\% 
\enddata
\end{deluxetable}

\section{Probabilities of strong lensing by nearby clusters}
In this section we will discuss how often one should expect to observe any signature 
of strong lensing effects in a sample of nearby clusters, taking into account 
some 
observational constraints. The lensing model adopted here assumes that the 
cluster mass
profile may be described by a singular isothermal sphere (SIS) and is
presented in detail in Appendix A. The model has 10 parameters. Each cluster is
characterized by two parameters: its redshift $z$ and the one-dimensional 
velocity
dispersion $\sigma_v$. The luminosity function of the field galaxies is described
by a Schechter function and has three parameters: $\phi^*$, $M^*$, and $\alpha$.
The analysis of the observations also has three parameters: the adopted limit
magnitude $m_l$, the seeing of the observations $\sigma_{seeing}$,  and the 
minimum  flux amplification by lensing, $A_{min}$. Note that in the SIS model 
for
strong lensing $A$ is also equal to the tangential
stretching of the arc or arclet. Once the cosmological model is
specified by the density parameters associated to the mass and to the
vacuum,  $\Omega_m$ and $\Omega_\Lambda$, all the parameters for the calculation 
are fixed and there is no dependence on the value of  $H_0$.

We adopt here the luminosity function derived in the Stromlo-APM 
Redshift Survey \citep{Lov92}, that is representative of the
local field 
galaxies. This luminosity function is well fitted by a Schechter function
with parameters $M^*_{b_j}=-19.50+5\log h$, $\alpha = -0.97$, and $\phi^*(0) =
1.40 \times 10^{-2} h^3$ Mpc$^{-3}$.
Assuming a color $(b_j - I) = 1.76$ \citep{Fuku95},
 that is 
appropriate for a Sbc galaxy (that is a good average of the local 
morphological mix
of galaxies), we have that $M^*_I=-22.8$ for 
$H_0 = 50$ km s$^{-1}$  Mpc$^{-1}$. For simplicity, we neglect any
evolution of the parameters $\alpha$ and $M^*$ with $z$.
This luminosity function is consistent with galaxy counts
in the I-band \citep{Smail95}.

\begin{figure}[h]
\epsfxsize=8cm
\centerline{\epsfbox{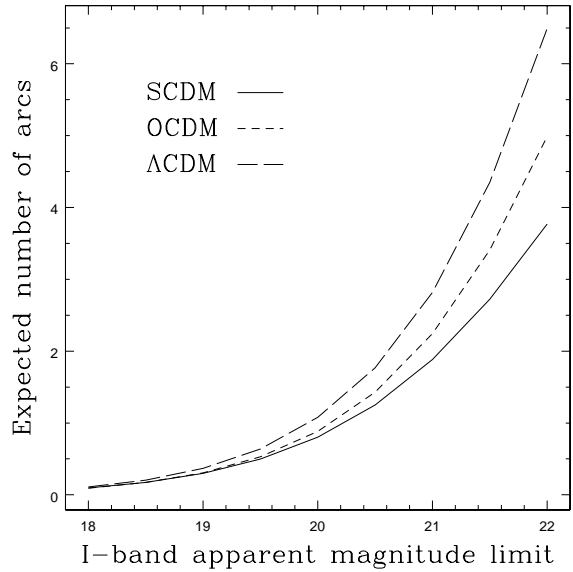}}
\vskip 0.5cm
\caption{Expected number of arcs in a cluster at $z=0.05$ and with velocity
dispersion $\sigma_v=1000$ km s$^{-1}$ as a function of the limit
magnitude in the I-band, when seeing effects are neglected. 
Results for three cosmological models are shown.}
\label{narc}
\end{figure}

We first discuss how the expected number of arcs varies with the magnitude 
limit
of the observations, neglecting seeing effects.
Figure \ref{narc} shows how the expected number of arcs with A$_{min}$ =2 
(the minimum
amplification produced by a SIS, see Appendix A), 
for a single cluster at $z=0.05$
with $\sigma_v = 1000$ km s$^{-1}$ (a massive, Coma-like cluster), varies
with the magnitude limit $m_l$ of the arc search. The results are
presented for three cosmological models: standard CDM (SCDM: $\Omega_M=1$, 
$\Omega_\Lambda=0$); open CDM (OCDM: $\Omega_M=0.3$,
$\Omega_\Lambda=0$); and
cosmological constant CDM ($\Lambda$CDM: $\Omega_M=0.3$,
$\Omega_\Lambda=0.7$). With $m_{l,I} = 19.0$ (the
limit adopted in the search), we have that $<N>$ is equal to 0.30, 0.31
and 0.37, for SCDM, OCDM and $\Lambda$CDM, respectively. A value of
$<N> = 1$ is achieved at $m_{l,I}= 20.2, 20.1$, and 19.9, for these
3 cosmologies. 
Figure \ref{narc} reveals that the expected number of arcs increases
strongly with  $m_l$, and searches going one magnitude deeper than ours may
plausibly find 2 to 3 times more evidence of strong lensing than our own
search. These results also indicate that different cosmological
models lead to similar results, at least for bright values of $m_l$. 
A major source of uncertainty in this 
kind of calculation
is due to the normalization of the luminosity function, $\phi^*$,
as evidenced by galaxy number counts in different directions.
Assuming an uncertainty of a factor 2 in $\phi^*$, $<N> = 1$ would be 
attained for $19.3 \simless m_{l,I} \simless 21.0$, where the brighter and
the fainter limits are for $\Lambda$CDM and SCDM, respectively.

\begin{figure}[h]
\epsfxsize=8cm
\centerline{\epsfbox{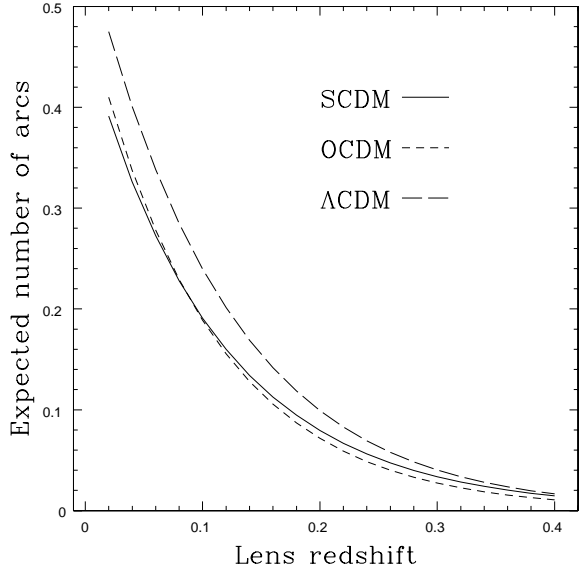}}
\vskip 0.5cm
\caption{Expected number of arcs in a cluster with velocity
dispersion $\sigma_v=1000$ km s$^{-1}$ as a function of its redshift, neglecting seeing
effects. 
Results for three cosmological models are shown.}
\label{znarcs}
\end{figure}

Figure \ref{znarcs} shows the dependence with the redshift of the expected 
number of arcs
with A$_{min}$ =2, for a cluster with $\sigma_v = 1000$ km s$^{-1}$, 
adopting m$_l$ as 19.0, for the same three cosmologies. Here too we are
neglecting seeing effects. This figure indicates that nearby clusters are 
more efficient  than far ones to produce arcs brighter than some
magnitude limit. This is due to the fact that low
redshift clusters project larger angular cross sections on to the plane of the 
sky than
more distant clusters.
This result is not in disagreement with the statistics of arcs as a function
of the lens redshift, which has a maximum in the range 0.2 $< z < $ 0.4 (e.g. 
\citep[e.g.][]{F&M94}. Indeed, Figure \ref{znarcs} shows the expected number of 
arcs per cluster, and
those arc statistics not only consider the lensing efficiency of a cluster at 
a given 
redshift, but also the total number of clusters (or the comoving volume if we 
consider a 
constant density of clusters) per bin of redshift too.
 
To calculate $<N>$ for our whole cluster sample, we need the velocity 
dispersion of the clusters. Unfortunately this quantity is not known for
5 clusters in the sample.  For A3408 we used
the value of \citet{CKH98} ``dark halo'' scenario.
The remaining 4 clusters were removed from this analysis (although they
have been included in the arc search described in the next section). 
Since their velocity dispersions are probably small (due to their low
richness and absence of detectable X-ray emission), their impact on the
results should be negligible. The values of $\sigma_v$ adopted
in the calculation were corrected to their rest frame values
(i.e., the actual value is the observed value that appears in Table 
\ref{sample} times $(1+z)^{-1}$).

Assuming $A_{min}$ =2 and the same  magnitude limit
adopted in the arc search, $m_{l,I}=19.0$, the expected number of arcs and 
arclets
in our cluster sample is 2.9 for SCDM, 3.0 for OCDM, 
 and 3.6 for $\Lambda$CDM.  
Seven clusters with $\sigma_v \ge 900$ km s$^{-1}$ contribute with
more than 55\% for $<N>$. Note that these are upper limits, actually, since 
this
estimate does not include the seeing.

The seeing may dramatically affect this kind of estimate, because it tends to 
circularize non-circular sources. In order to quantify seeing effects, we 
have made 
simulations, using the IRAF package {\it artdata}. We have simulated  
``arcs'' as exponential profiles with central surface brightness
19.9 I-band mag arcsec$^{-2}$ \citep[typical of a][galaxy disc]{Free70}
 with several total apparent magnitudes and axial ratios, assuming the same 
 sky level
and noise of our images. These simulated  images have been convolved with a 
Moffat 
PSF with FWHM  1$^{\prime\prime}$.4
(the average seeing of our images) and then their magnitudes and
axial ratios were measured with the software SExtractor \citep{B&A96},
using the same parameters used in the search for arcs (Section 4).

\begin{figure}[h]
\epsfxsize=8cm
\centerline{\epsfbox{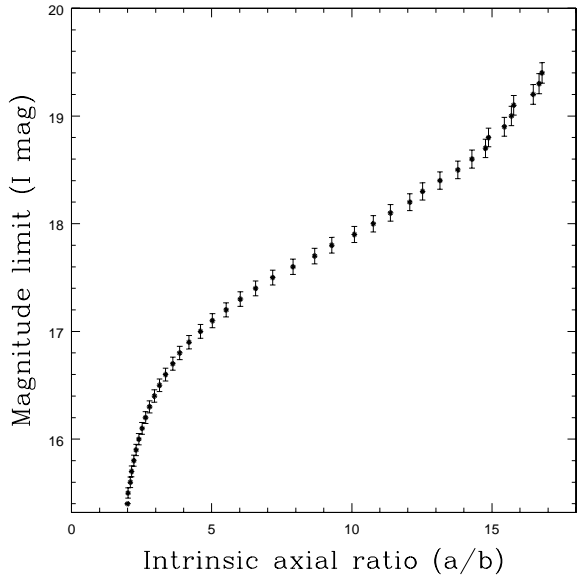}}
\vskip 0.5cm
\caption{Magnitude limit as a function of the tangential stretching (or
amplification) obtained using the average of 500 simulations per axial ratio
with 
{\it artdata}. The error bars correspond to 1 $\sigma$ statistical uncertainity.
The seeing FWHM 
adopted is 1.4$^{\prime\prime}$, the average of the observations.}
\label{amp}
\end{figure}

Let us assume that we can identify a gravitational arc if it has an axial 
ratio of at least
1.5. For a given value of its intrinsic axial ratio, this arc would be 
identified only if
its magnitude is lower than some limit $m_{max}$, because fainter images would 
appear
with amplification inferior to 1.5 due to the seeing. Large and bright arcs 
are not
strongly affected by the seeing, 
but faint and small arclets are.
With 500 simulations for 50 values of
the intrinsic axial ratio, we have estimated the mean value of $m_{max}$. 
The results are illustrated in Figure 4. 
Another result of the simulations is that the
difference between total and isophotal magnitudes near the limit of detection
is about 0.5. Considering that the observed limit of completness is $\sim$ 19.5,
 we conclude that the total magnitude limit is $\sim$ 19.0 mag.    

\begin{figure}[h]
\epsfxsize=8cm
\centerline{\epsfbox{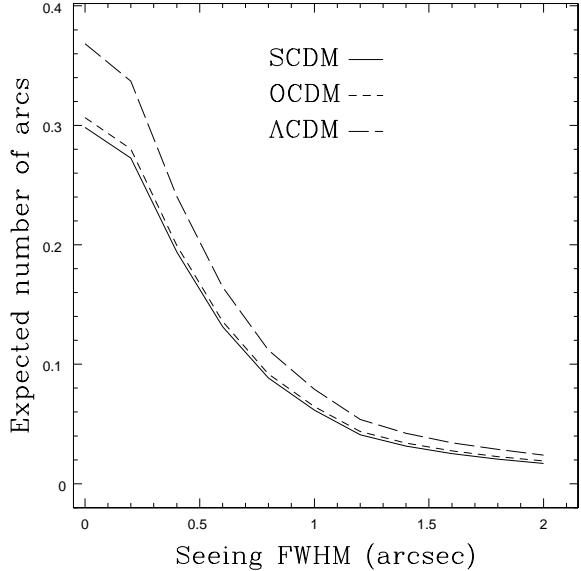}}
\vskip 0.5cm
\caption{Expected number of arcs in a cluster with velocity
dispersion $\sigma_v=1000$ km s$^{-1}$ at $z=0.05$ and m$_l$ = 19.0,
as a function of
the seeing (FWHM). Results for three cosmological models are shown.}
\label{seeing}
\end{figure}

We present in Figure \ref{seeing} the expected number of arcs for a cluster 
with
$\sigma_v = 1000$ km s$^{-1}$ at $z=0.05$. This figure shows how  dramatic
are seeing effects for lensing detection. The results displayed in Figure 
\ref{seeing} 
for each cosmological model may be parametrized as 
$N_0 \exp{(\beta \sigma_{seeing})}$, 
where $N_0$ is the expected number of arcs in the absence of image
degradation due to the seeing. A value of $\beta = -1.5$ provides a good fit 
to the data.

We have recalculated the estimates of the number of arcs in our sample taking 
the seeing into account. The expected values are
0.32, 0.34, and 0.42, for SCDM, OCDM and $\Lambda$CDM models, respectively. 
Note that the inclusion of the seeing in the analysis has lead to a
substantial decrease in the number of expected arcs; 
 these numbers are $\sim$9 times smaller than when seeing effects are
neglected. 
Using the OCDM results and  Poissonian statistics, the probability of one arc 
detection in the sample is 24.2\%, two detections is 4.1\%, three detections is
0.5\%, and no detections is 71.2\%. Hence, our estimation is more consistent
with no detections, but the probability of finding at least one arc is not 
negligible.

\section{Search of bright arcs and arclets}

In this section we describe the procedure we have adopted in the search for
arcs  and arclets produced by gravitational lensing in the sample of clusters
presented in Section 2. 
We have looked for evidence of strong lensing not only
in the regions corresponding to the central parts of the clusters, where
the projected mass density is high and, consequently, the probability of
lensing is higher than in other regions, but also around bright galaxies.

Our strategy for the search was the following. Initially, we made a catalog
of galaxies using SExtractor. We
adopted a detection threshold of 1.5 $\sigma$ over the sky level (which 
corresponds to $I \sim$ 23.3 mag arcsec $^{-2}$) and a minimum detection 
area of 40 pixels (or 6.27 arcsec$^2$). The distribution of magnitudes
of the galaxy catalog presents a cutoff at $I \sim 20$, indicating that 
its completeness limit is at $I \sim 19.5$. 

Afterwards, all galaxy images with semi-major axis larger
than 8 pixels (3.17 arcsec) were modeled with the STSDAS/Ellipse package,
and the model images were subtracted from the actual galaxy images. The
aim here was to reveal any arc-like structure superposed on to a galaxy image.
In general this procedure worked well, showing the presence of several
objects within the galaxy image. This procedure tends to produce a central
residual, due to the seeing and pixelization but, since these
residuals are usually
restricted to the very central regions of a galaxy image, they do not have
a relevant impact in our arc survey.
During the process of image subtraction we inspected by eye most of the
images in each field, paying special attention to objects with axial
ratio larger or equal to 1.5. 

After the subtraction of the galaxy images, we created a new SExtractor
galaxy catalog and visually inspected all new objects contained
in this second catalog, focusing again on the more elongated ones.
At first we selected about twenty candidates,  elongated
objects which could not be morphologically identified, by visual inspection,
as edge-on spirals. Then we verified whether these objects were 
tangentially or radially 
disposed with respect to the cluster center or some bright, nearby galaxy. 
After this stage, only two candidates remained.
The first is the same arclet discovered by \citet{CeH}
in A3408 and discussed by  \citet{CKH98}. The
second candidate was found in the cluster A3266. It is not in
the central region of the cluster but near
one bright elliptical galaxy. It is, hence, a candidate for 
lensing by a cluster substructure, instead of lensing by the cluster overall 
potential.

\section{Discussion}
In this section we discuss the main characteristics of the two arclet
candidates, taking into account some new follow-up observations.

\subsection{The arclet in A3266}
The cluster A3266 (also known as S\'ersic 40/6) is apparently regular, 
with type I-II in the Bautz-Morgan system. Its center contains a very tight 
dumbbell 
pair, at $\alpha = 4^h 31^m 14.25^s$ and 
$\delta = -61^\circ 27^\prime 11.3^{\prime \prime}$ (J2000). 
The recession velocity of the cluster relative to the CMBR is 17782 km s$^{-1}$.
A detailed analysis of this
cluster by \citet{QRW96} reveals that it has a large velocity
dispersion, 1306 $\pm$ 73 km s$^{-1}$ within $\sim 1 h^{-1}$ Mpc, that has been 
interpreted as an evidence that A3266 is indeed the result of a recent
merger of two structures of comparable masses. This interpretation is also 
supported
by numerical simulations \citep{FQW99} as well as by an analysis
of the X-ray brightness distribution \citep{MFG93}. 

\begin{figure}[h]
\epsfxsize=16cm
\centerline{\epsfbox{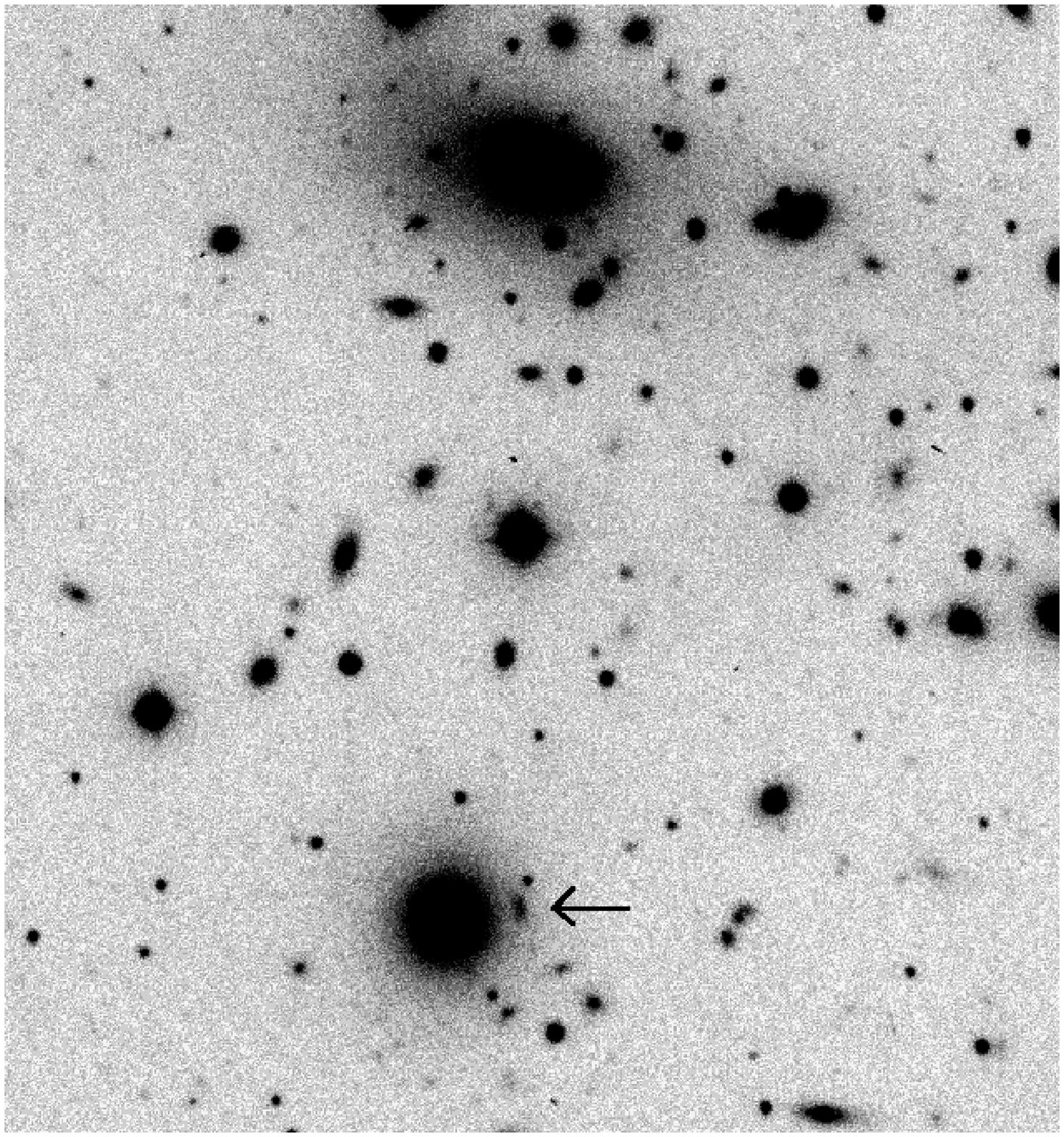}}
\vskip 0.5cm
\caption{Putative arc (indicated by an arrow) in Abell 3266. The bright
elliptical galaxy aside of the arc is the second brightest galaxy 
(apart from the central dumbbell pair) of this
cluster in the I-band and is 2.9 arcminutes distant from the cluster 
center seen at the top
of the figure. North is at the top and East is at the left.}
\label{arc}
\end{figure}

We present in Figure \ref{arc} part of the I-band image of A3266, taken with 
the 
0.9m CTIO telescope. The candidate arclet is indicated with an arrow. 
Its centroid is at $\alpha = 4^h 31^m 15.53^s$ 
and $\delta = -61^\circ 30^\prime 3.7^{\prime \prime}$ (J2000).
It is at 16.6 arcsec (29 h$_{50}^{-1}$ kpc at the cluster distance) from the 
center 
of a nearby, bright elliptical, and 
at 2.89 arcminutes (303 h$_{50}^{-1}$ kpc) from the center of the cluster (at 
the position of the dumbbell pair).
At the isophotal level of  23.3 mag  arcsec$^{-2}$ (1.5$\sigma$ over the sky 
background) its magnitude, semi-major axis, axial ratio, and position angle  
are 
$I = 18.89 \pm 0.03$, $a = 4^{\prime\prime}.1$, $a/b = 2.11
\pm 0.34 $, and $\theta = 3.5^\circ \pm 2.9^\circ$, 
respectively.

The elliptical galaxy near the object is located at $\alpha = 4^h 31^m 16.6^s$ 
and $\delta = -61^\circ 30^\prime 8^{\prime \prime}$ (J2000) and is the 
second  brightest galaxy of the cluster in the I-band (apart from the central 
dumbbell pair).
Its heliocentric radial velocity is 
15819 $\pm$ 30 km s$^{-1}$ and its apparent total B magnitude is 15.40 $\pm$ 
0.20
\citep{deVau91}, corresponding to an absolute magnitude of -22.38. 

Spectroscopic observations of the arclet by W. Kunkel (private communication)
with the 2.5m Dupont telescope have shown that this object has an heliocentric
radial velocity of 21900 km s$^{-1}$.
With this velocity, this object may be inside the cluster, since most of the 
cluster members have velocities between 15000 km s$^{-1}$ and 21000 km s$^{-1}$ 
\citep{QRW96}
On the other hand it can also be behind the cluster, which strengths the 
probability  that it has been lensed either
by the cluster potential or by the nearby galaxy, or both. 
In particular we want to test the hypothesis that this is a background galaxy
being lensed by a mass peak associated with the bright elliptical near the arc.

Let us assume a SIS model for the mass distribution centered at the center of the
nearby, bright elliptical galaxy. From the observed elongation of the arc candidate
(that is equal to its amplification), the distance between the center of the lens 
and the arc image is about two critical radius (see Appendix, Equation \ref{er}).
Hence, the SIS should have a velocity dispersion of 1265 km s$^{-1}$, 
 approximatelly the same of  
the cluster itself. The mass enclosed within the arc radius 
is $3.0 \times 10^{13}$ h$_{50}^{-1}$ M$_\odot$.
The bright galaxy near the arc has $M_I = -24.65$ at the 23.7 mag
arcsec$^{-2}$ level. Assuming $V-I = 1.31$, appropriate
for an elliptical galaxy \citep{Fuku95} the
arc would be explained by our lens model if 
M/L$_V$=163 h$_{50}$ M$_\odot$/L$_\odot$ for this system.
Such a high M/L value could be expected if this galaxy were at the center of a 
massive sub-structure. However, neither the X-ray map of the 
cluster \citep{J&F99} nor the  dynamical analysis of 
\citet{QRW96} present any evidence of significant sub-structure
at this position. Possibly, this arc-like feature is a disk galaxy
(or the bar of a disk galaxy) that is a cluster member (or is not far from it)
instead of a real arclet.

\subsection{The arclet in A3408}
This structure has been discussed by \citet{CKH98},
 who successfully modeled it as a galaxy at $z=0.073$ lensed
by the cluster A3408, at $z=0.042$. The adopted cluster mass distribution
is a scaled version of mass profiles derived from the study
of high redshift clusters. Their preferred model has a component 
that follows the brightness profile of the central elliptical galaxy
and is immersed in a massive dark halo. From the lens model and the
equivalent widths of some prominent emission lines ([O II] $\lambda$3727,
[O III] $\lambda$5007 and H$\alpha$) they have suggested that the source galaxy
is probably a spiral with intrinsic diameter 14.6 kpc and magnitude $M_B=-18.2$.

We have imaged the central part of A3408 with the 0.9m telescope of CTIO with
interference filters with the aim of detecting other galaxies at the redshift
of the source, which could help to improve the lens model. We have used two
different filters. One is centered at 7053 \AA, with FWHM of 79 \AA, 
which allows the detection of H$\alpha$ at the redshift of
the source, covering a velocity range of 1700 km/s. The other is centered at 
6961 \AA,  with FWHM of 79 \AA, and samples the continuum near
the H$\alpha$ line. We have used the Tek2k3 CCD to make 5 images of 15 minutes each 
in each of the two filters.

\begin{figure}[h]
\centerline{{\epsfxsize=8cm \epsfbox{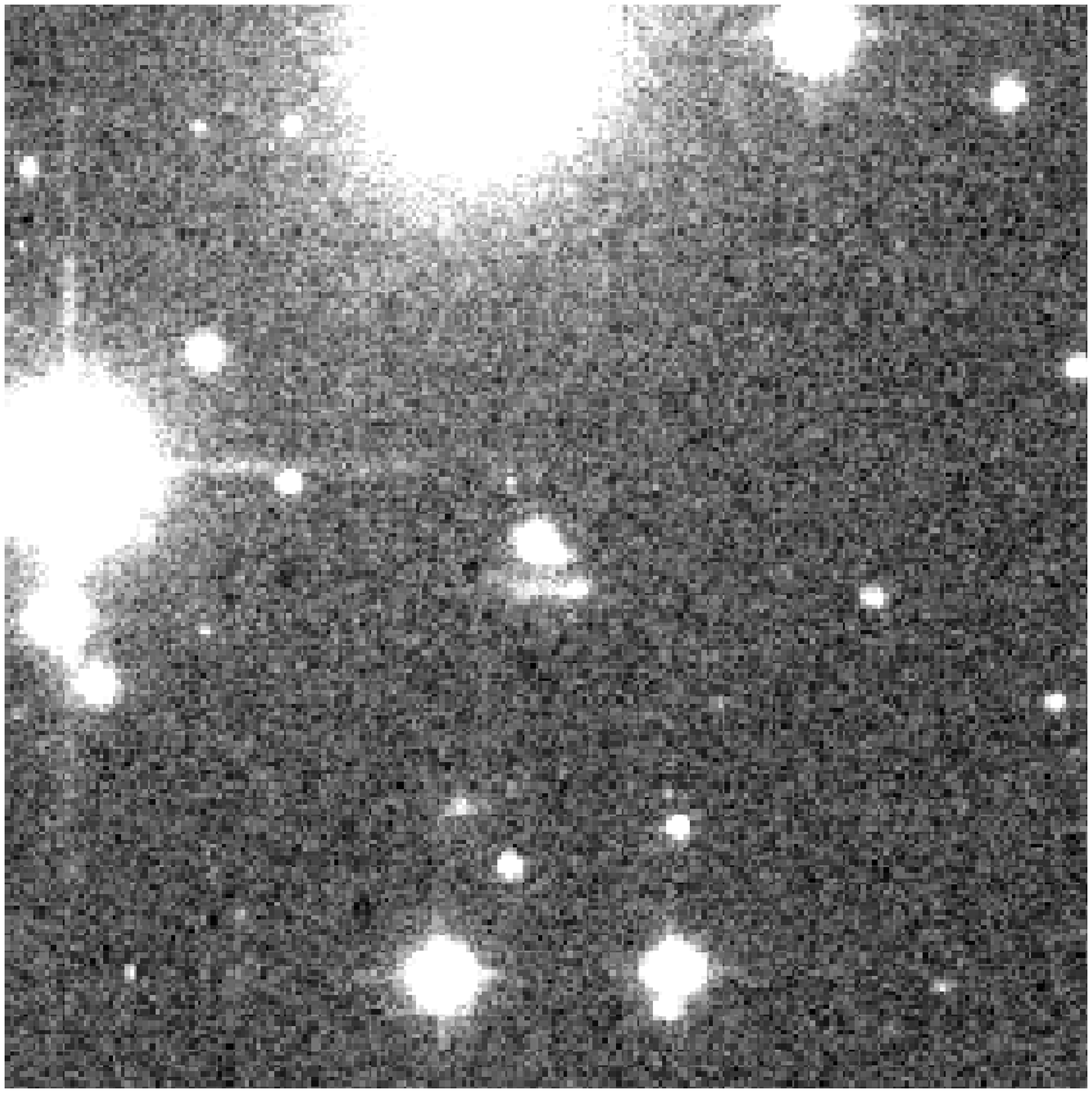}} 
{\epsfxsize=8cm \epsfbox{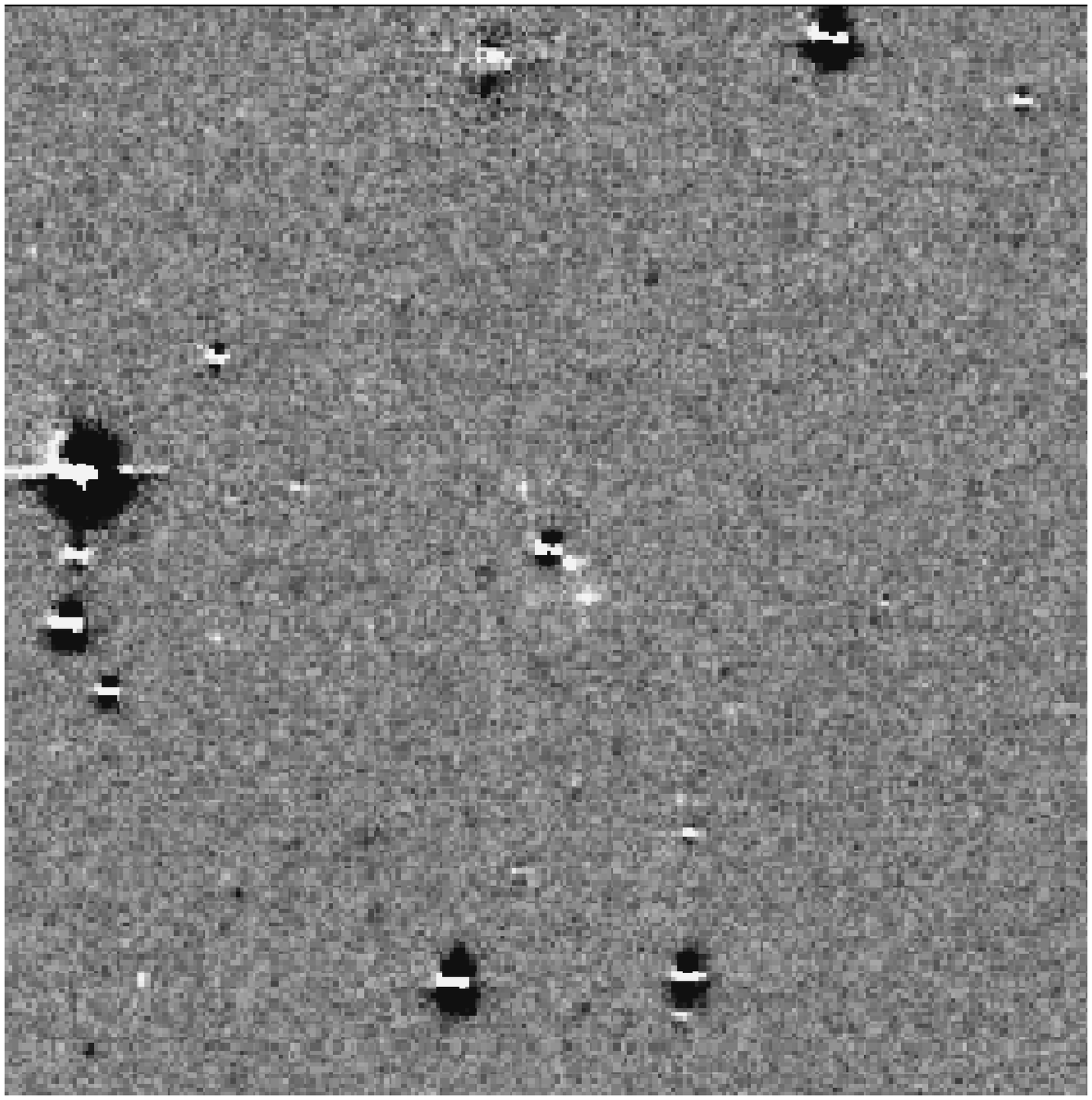}}}
\caption{Image of the gravitational arc in A3408 in  H$\alpha$ (left), and
the residuals of the subtraction between the images taken in the H$\alpha$ line 
and in the adjacent continuum (right). North is at the top and East is at 
the left.}
\label{ha}
\end{figure}

The images were reduced using standard procedures with IRAF. They were stacked 
and normalized so that at the end of the reduction we had two images (one for
each filter) where the mean flux (counts) of the stars were the same in both 
images. After that, we produced a new image by subtracting the image taken
with the continuum filter from the image taken with the filter centered in 
H$\alpha$. Figure \ref{ha} presents both the H$\alpha$ and the residual
images of the central region of A3408, centered at the position of the star 
near the arc position. The image containing the residuals indicates that the
image subtraction was good, despite the features that remained at the center
of the star and galaxy images (produced by seeing and pixelization effects), 
since most of the extended regions of the object images were removed. 
Figure \ref{ha} also indicates that the H$\alpha$ emission of the arc is not
uniform and is strongest at the western side of this object.

\begin{figure}[h]
\epsfxsize=8cm
\centerline{\epsfbox{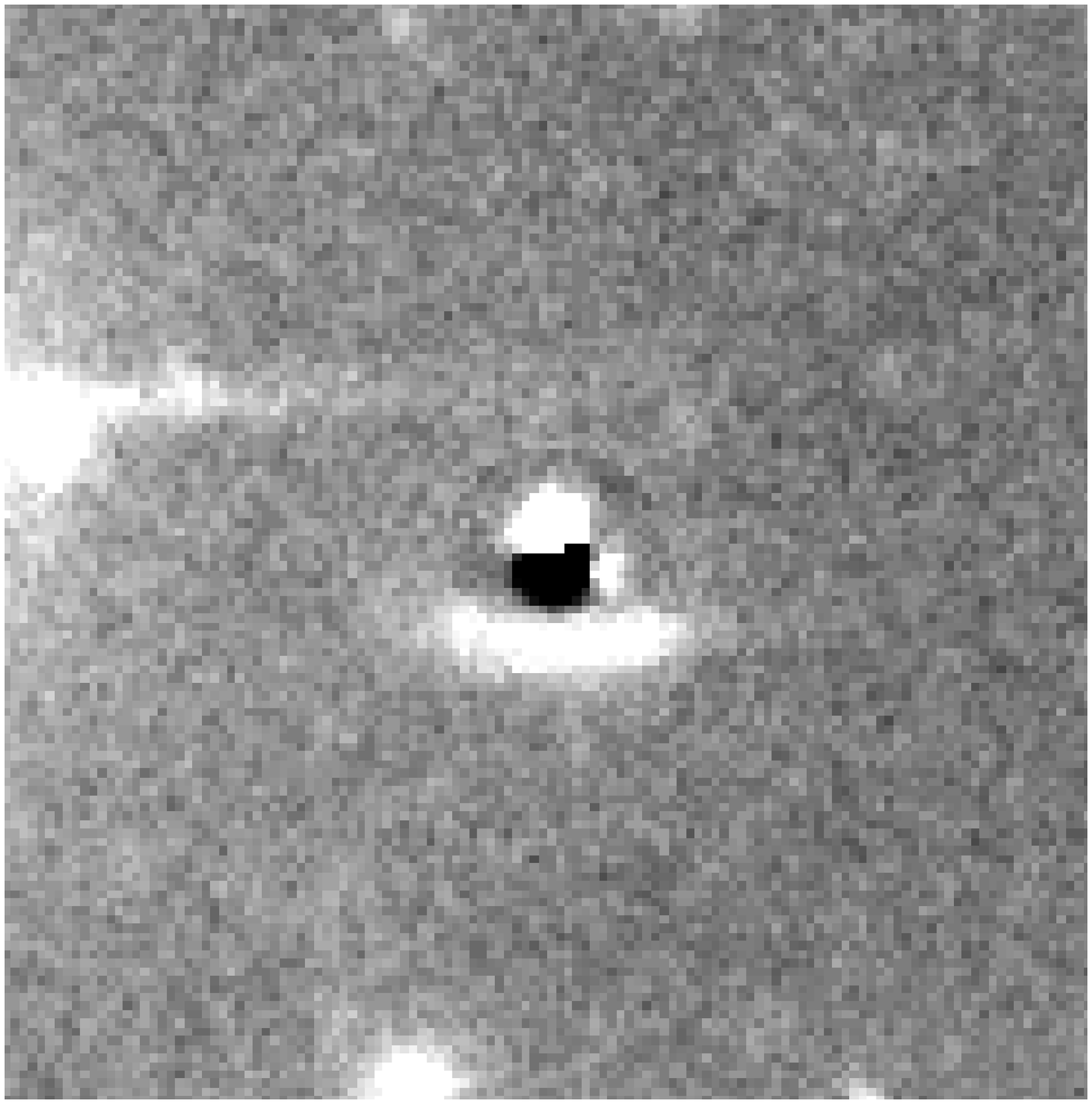}}
\vskip 0.5cm
\caption{Broad band image (B+R+I) of the arc in Abell 3408, with the image of
the star near the arc subtracted.}
\label{point}
\end{figure}

An interesting feature present in the residual image is a point-like object,
between the side of the arc with the strongest  H$\alpha$ emission and the
star below the arc. This object  also appears in broad band images, after
removing the image of this star, as shown in Figure \ref{point}. This
feature is probably a companion galaxy of the arc source, at approximately
the same redshift. Unfortunately, the presence of the star precludes
further analysis of this object and its use to constrain more sophisticated
lens models. No other objects are seen in Figure  \ref{ha} at the same
redshift of the arc. Note that this does not mean that a galaxy group (containing
the arc source) can not exist, because only their brightest members could
be detected in these images. Moreover the CCD covers an area of 1.52 $\times$
1.52 h$^{-1}_{50}$Mpc at z = 0.073. Some loose groups occupy areas larger than it 
(up to 5 Mpc of side), so we might not be sampling the entire group; 
however, in this case the over-density caused by this 
group behind A3408 might be not so significant and the impact of its presence
on our probability calculation would be small.

Overall, A3408 seems to be a an interesting structure. It is a poor cluster, 
with low X-ray emission but  high lensing mass, consistent with its
velocity dispersion of  900 km s$^{-1}$. On the other side, it may be 
in interaction with A3407, since they are
very close to each other \citep{G93}.
Weak lensing observations of this 
cluster, combined with other methods of mass determination, like dynamical 
or X-ray analysis, should give us a better understanding of  this system. 

\subsection{Comparison of theoretical and observational results}

As discussed in Section 3, the probabilities for 0 and 1 detection of
a gravitational arc in the cluster sample discussed here
and including seeing effects  are 71\% and 24\%,
respectively. On the other side, our search for arcs in this sample 
produced two candidates, one in A3266 and the other in A3408. The 
arclet in A3408 is indeed good evidence of gravitational lensing by
a nearby cluster \citep{CKH98}. The arc-like object
in A3266, however, is more difficult to be
interpreted as a result of gravitational lens distortion, because it
would require a very massive substructure around the bright 
elliptical galaxy near the object. Note that one arc detection is
consistent with our theoretical expectations but, of course, this
does not allow us to strongly constrain any of the model parameters.

It is worth stressing the relative role played by the seeing and
the magnitude limit in arc searches. As shown in Section 3, the 
expected number of arcs increases fast with the limiting magnitude 
of a survey, but seeing strongly affects fainter galaxies. Better
seeing conditions lead to a substantial increase in the efficiency of these 
surveys. Searches of strong lensing effects in clusters will greatly
benefit of the improvement in image quality that is arriving with
active and adaptive optics cameras and new generation telescopes.

It is interesting to compare our results with those of \citet{Lup99}.
They have searched for arcs and arclets in a sample of 38 X-ray selected 
clusters ($L_X \ge 2\times 10^{44}$ erg s$^{-1}$) from the {\it Einstein Observatory 
Medium Sensitivity Survey} (EMSS) in the redshift range 0.15 $\le z \le$ 0.823.
Their images were obtained with the University of Hawaii 2.2m telescope in the
R band for all clusters, and in the B band for most of them. The median seeing 
in R was 0.8$^{\prime\prime}$.
They found arcs and arclets in 8 clusters, or 21\% of the sample, that can 
be compared with our success rate of only 3\% (1 arc in 33 clusters). This 
discrepancy may be explained by taking into account
that the two samples are very different regarding redshift distribution and
image quality (resolution and magnitude limit).
 The images of 
Luppino et al. were obtained under better seeing conditions and are substantially
deeper than ours. Besides, the EMSS sample, based on X-ray luminosities, 
is biased towards massive clusters.

\section{Summary}
We have discussed strong lensing effects produced by nearby clusters of galaxies.
Using a simple mass model for the clusters we have shown that the expected number
of arcs or arclets expected in a sample of nearby clusters is not as small as
is usually thought. The results are strongly dependent on
the magnitude limit and the seeing quality of the imaging. 

Our search for arcs and arclets  in a  sample of 33 nearby clusters has resulted in two
arc candidates, one in A3408 and the other in A3266. The first was already discovered 
by \citet{CeH} and modeled by \citet{CKH98}. Observations
in H$\alpha$ reported here show the presence of another object near the arc at approximately
the same redshift, but contamination from a foreground star precludes its
further use  to constrain the
cluster mass model. Our analysis of the arclet candidate in A3266 suggests or
a false detection or that the mass concentration necessary 
to explain this structure by lensing is very strong.

Our result indicates that deep imaging of nearby clusters under good seeing
conditions
may be extremely useful for high-resolution 
mapping of the mass distribution of these structures, since the probability of
detection of strong lensing features will be enhanced. Additionally, deep imaging
with mosaic CCD detectors \citep[e.g.][]{Jetal99}
also allow the detection of weak lensing, which may
provide more constraints on the mass distribution in the central regions of 
the galaxy clusters.

\acknowledgments
ESC and LS gratefully acknowledges support by Brazilian agencies FAPESP,
CNPq and PRONEX . RG acknowledges support by NSF grant AST-9617069.

\appendix
\section{Estimation of the expected number of gravitational arcs and arclets in
a sample of clusters}
We assume here that the matter distribution of a galaxy cluster can be described by a
singular isothermal sphere (SIS). The reason for this choice is twofold. First,
this is the simplest model, with only one parameter: the one-dimensional velocity
dispersion $\sigma_v$. Second, in this model one can easily relate amplification of a
background source and its tangential stretch to the impact parameter (angular
distance between the source and the center of the lens). We also assume that
galaxies follow dark matter and have the same velocity dispersion.

More complex models may not be
necessary, given the uncertainties in many parameters that enters in this
calculation.  For instance,
if instead of a SIS we have considered a model with a core radius, the
lensing probability would decrease \citep[e.g.][]{W&H93}.
Conversely, an ellipticity in the galaxy distribution, as well as
the presence of sub-structures in a cluster, increases the lensing
probability, due to enhanced tidal effects introduced by the asymmetry
in the mass distribution \citep{Bar95}. Hence, the effects due to
the inclusion of a core radius and cluster ellipticity may more or less
cancel each other. Note also that the determination of these two quantities
is not easy, being very sensitive to the choice of the cluster center and
the presence of sub-clustering. Of course, even more refined mass
models of each cluster are possible, where individual galaxy halos are
taken into account \citep[e.g.][]{K96,Nat98,G&S99,Bez99}, but such
models are beyond the scope of this simple calculation. Anyway, we do not
expect that the order of magnitude of the results presented here
will change dramatically with the use of more sophisticated mass models.

Arcs and arclets \citep[see][for a review]{F&M94} are a result of
strong lensing by galaxy clusters of extended sources close to cusps or
higher order catastrophes in the source plane. In what follows we will
assume that an arc or arclet is produced if a source bright enough falls
within the critical circle of the cluster in the source plane.
It is well known that a spherically symmetric lens like our SIS model
does not produce cusps. However, this drawback does not precludes the use
of this sort of model to obtain estimates of the the strength of lensing
effects. Indeed, such an approach has been extensively applied to
compute  lensing cross sections \citep[e.g.][]{B93,CQM99,C99}.

For a singular isothermal sphere be able to act as a strong lens, 
the source (in the source plane) should be within the
critical circle inside which the mean mass surface density of the cluster is
greater than a certain critical density that depends only of the relative
distances between observer, cluster, and source (and hence of the cosmology).
The angular radius of this critical circle (centered in the cluster center)
is
\eq
\theta_c =  4 \pi (\sigma_v/c)^2{D(z_l,z_s) \over D(0,z_s)}
\eeq
where 
$D(z_l,z_s)$ and $D(0,z_s)$ are the angular diameter distances between the 
lens (at redshift $z_l$) and the source (at $z_s$), and between the observer 
($z=0$) and the source, respectively. The diameter distances are computed
adopting the analytical filled-beam approximation \citep{Fuku92}.
Following \citet{B&K87}, the magnification of the brighter of 
the two images produced by this lens is given by
\eq
\label{er}
A = 1 + {\theta_c \over \theta_s}
\eeq
where $\theta_s$ is the angular distance on the plane of the sky 
between the center of the lens and the source (the impact parameter).

The expected number of arcs with magnification larger than a certain value 
$A_{min}$ (the minimum amplification produced by a SIS is 2) due to
a cluster at redshift $z_l$ can be written as
\eq
< N(z_l) > = \int_{A_{min}}^\infty dA \int_{z_l}^\infty dz~ N(z,A)
\eeq
where $N(z,A) dA dz$ is the number of observable sources between $z$ and
$z+dz$ that suffer magnifications between $A$ and $A+dA$ by the 
gravitational field of the cluster. These are the sources in that redshift
interval with luminosities larger than $L_{min}(z)/A$, where $L_{min}(z)$
is the minimum luminosity that a source at redshift $z$ should have to
be included in the sample (in the absence of lensing effects), 
and that are inside a ring with solid angle
$d \Omega = 2 \pi \theta d\theta$. Here $\theta$ is the angular distance 
to the cluster center corresponding to a magnification $A$ of the source
luminosity, accordingly to Equation (\ref{er}). It is easy to verify that
\eq
d\Omega = 2 \pi \theta_c^2
{ d A \over (A-1)^3}
\eeq

We can also write
\eq
\label{N}
N(z,A) dA dz = n(z) dV(z) {d\Omega \over 4 \pi}
\eeq
where $n(z)$ is the mean number density of the sources at redshift $z$
that are bright enough to be detected:
\eq
n(z) = \int_{L_{min}(z)/A}^\infty \phi(L,z) dL
\eeq
where the term $L_{min}(z)/A$ takes into account the magnification bias,
by which some sources are magnified by lensing, acquiring an apparent
luminosity high enough to be included in a magnitude limited sample.

The differential luminosity function at $z$, $\phi(L,z)$, may be described
by a Schechter function in which the comoving density of galaxies at
redshift $z$ with luminosities between $L$ and $L+dL$ is
\eq
\phi(L,z) dL = \phi^*(z) \left({L \over L^*(z)}\right)^{\alpha (z)}
\exp \left(-{L \over L^*(z)}\right) {dL \over L^*(z)}, 
\eeq
where $\phi^*(z)$, $L^* (z)$ and $\alpha (z)$ are the parameters of
the luminosity function at redshift $z$.
Then,
\eq
n(z) = \phi^*(z) \Gamma\left[{1+\alpha(z), {L_{min}(z,m_l) \over A~L^*(z)}}\right]
\eeq
Assuming that the number of sources in a comoving volume is conserved,
we have that
\eq
\phi^*(z) = (1+z)^3 \phi^*(0)
\eeq
where $\phi^*(0)$ is the normalization of the local ($z=0$) luminosity 
function.

The luminosity $L_{min}$ is related to the magnitude limit $m_l$
considered in the analysis and the magnification $A$ by
\eq
\label{lum}
{L_{min}(z,m_l) \over L^*(z)} = A \times 10^{-0.4(m_l - 
5 \log[(1+z)^2 D(0,z)] -25 - M^*)}
\eeq
where $z$ is the redshift of a source and $M^*$ is the local
value of the characteristic magnitude of the luminosity function in
the photometric band of interest.
Note that we are assuming that the source galaxies and $M^*$ have the same 
$k$ and evolutive corrections. 

The volume element $dV(z)$ that also appears in Equation (\ref{N})
is the proper volume between $z$ and $z+dz$ and may be written as
\eq
{dV(z) \over dz} = {4 \pi c D^2(0,z) \over H_0} 
{1 \over (1+z) E(z)},
\eeq
where 
\eq
E(z) = [(1+z)^3 \Omega_M+(1+z)^2 (1-\Omega_M-\Omega_\Lambda)
+\Omega_\Lambda]^{1/2}
\eeq
and where $\Omega_M$ and $\Omega_\Lambda$ are the density parameters for matter
and vacuum energy, respectively.

Putting all together, the expected number of arcs produced by a cluster
at redshift $z_l$ is

\begin{eqnarray}
\label{narcold}
&&<N(z_l,\sigma_v)>  ~ =  {32 \pi^3 \phi^*(0) \sigma_v^4 \over c^3 H_0} \times \nonumber \\ [0.4cm]
&& \times
\int_{A_{min}}^\infty {dA \over (A-1)^3}  
 \int_{z_l}^{z_{max}} dz
{(1+z)^2 D(z_l,z)^2 \Gamma\left[{1+\alpha(z), 
{L_{min}(z,m(A)) \over A~L^*(z)}} \right] \over E(z)}
\end{eqnarray}

where we have adopted $z_{max} = 3$ (our results are insensitive to 
this limit of integration, because the number 
of sources brighter than $m_l$ decreases quickly with $z$).

A more realistic model must, however, take into account the effect of the seeing,
that tends to circularize object images, specially the faint ones. The
simulations described in Section 3 show that, for a given intrinsic stretching 
(or amplification $A$), the observed axial ratio will be larger than 1.5 (the minimum
value adopted in the arc search discussed in Section 4) only for objects brighter
than a certain apparent magnitude $m_{max}(A)$. Hence, seeing effects may be
included in the analysis by using the minimum of $m_l$ and $m_{max}(A)$ instead of
$m_l$ in Equation (\ref{lum}). The relation between $m_{max}$ and $A$ adopted here is
shown if Figure \ref{amp}.

For a sample containing $N_c$ clusters, the total expected number of arcs
is given by
\eq
<N> = \sum_{i=1}^{N_c} <N(z_i,\sigma_{v,i})>
\eeq
where $z_i$ and $\sigma_{v,i}$ are the redshift and  velocity dispersion 
of the $i-$th cluster.

\end{document}